\documentstyle[12pt]{article}
\catcode`\@=11
\def\section{\@startsection {section}{1}{\z@}{3.ex plus 1ex minus
 .2ex}{2.ex plus .2ex}{\raggedright\large\bf}}
\def\subsection{\@startsection{subsection}{2}{\z@}{2.75ex plus 1ex minus
 .2ex}{1.5ex plus .2ex}{\raggedright\bf}}
\catcode`\@=12
\newskip\humongous \humongous=0pt plus 1000pt minus 1000pt

\newif\ifdtup

\def\oldreffmt#1{\rlap{[#1]} \hbox to 2\parindent{}}

\def\figfmt#1{\rlap{Figure {#1}} \hbox to 1in{}}

\def\beq{\begin{equation}}
\def\eeq{\end{equation}}
\def\bea{\begin{eqnarray}}

\def\com#1#2{
        \left[#1, #2\right]}
\def\eea{\end{eqnarray}}
\def\ap#1,#2,#3#4{           {\it Ann. Phys. (NY) }{\bf #1} (19#3#4) #2}
\def\apj#1,#2,#3#4{          {\it Astrophys. J. }{\bf #1} (19#3#4) #2}
\def\apjl#1,#2,#3#4{         {\it Astrophys. J. Lett. }{\bf #1} (19#3#4) #2}
\def\app#1,#2,#3#4{          {\it Acta Phys. Polon. }{\bf #1} (19#3#4) #2}
\def\com#1,#2,#3#4{          {\it Comm. Math. Phys. }{\bf #1} (19#3#4) #2}
\def\ib#1,#2,#3#4{           {\it ibid. }{\bf #1} (19#3#4) #2}
\def\nat#1,#2,#3#4{          {\it Nature (London) }{\bf #1} (19#3#4) #2}
\def\np#1,#2,#3#4{           {\it Nucl. Phys. }{\bf B#1} (19#3#4) #2}
\def\npps#1,#2,#3#4{  {\it Nucl. Phys. B (Proc. Suppl.) }{\bf B#1} (19#3#4) #2}
\def\pl#1,#2,#3#4{           {\it Phys. Lett. }{\bf #1B} (19#3#4) #2}
\def\pla#1,#2,#3#4{          {\it Phys. Lett. }{\bf #1A} (19#3#4) #2}
\def\pr#1,#2,#3#4{           {\it Phys. Rev. }{\bf D#1} (19#3#4) #2}
\def\prep#1,#2,#3#4{         {\it Phys. Rep. }{\bf #1} (19#3#4) #2}
\def\prl#1,#2,#3#4{          {\it Phys. Rev. Lett. }{\bf #1} (19#3#4) #2}
\def\pro#1,#2,#3#4{          {\it Prog. Theor. Phys. }{\bf #1} (19#3#4) #2}
\def\rmp#1,#2,#3#4{          {\it Rev. Mod. Phys. }{\bf #1} (19#3#4) #2}
\def\sp#1,#2,#3#4{           {\it Sov. Phys.-Usp.}{\bf #1} (19#3#4) #2}

\def\zp#1,#2,#3#4{           {\it Zeit. fur Physik }{\bf #1} (19#3#4) #2}
\catcode`\@=11
\def\eqnarray{\stepcounter{equation}\let\@currentlabel=\theequation
\global\@eqnswtrue
\global\@eqcnt\z@\tabskip\@centering\let\\=\@eqncr
\gdef\@@fix{}\def\eqno##1{\gdef\@@fix{##1}}%
$$\halign to \displaywidth\bgroup\@eqnsel\hskip\@centering
  $\displaystyle\tabskip\z@{##}$&\global\@eqcnt\@ne
  \hskip 2\arraycolsep \hfil${##}$\hfil
  &\global\@eqcnt\tw@ \hskip 2\arraycolsep $\displaystyle\tabskip\z@{##}$\hfil
   \tabskip\@centering&\llap{##}\tabskip\z@\cr}

\def\@@eqncr{\let\@tempa\relax
    \ifcase\@eqcnt \def\@tempa{& & &}\or \def\@tempa{& &}
      \else \def\@tempa{&}\fi
     \@tempa \if@eqnsw\@eqnnum\stepcounter{equation}\else\@@fix\gdef\@@fix{}\fi
     \global\@eqnswtrue\global\@eqcnt\z@\cr}

\newcount\hour
\newcount\minute
\newtoks\amorpm
\hour=\time\divide\hour by 60
\minute=\time{\multiply\hour by 60 \global\advance\minute by-\hour}
\edef\standardtime{{\ifnum\hour<12 \global\amorpm={am}%
	\else\global\amorpm={pm}\advance\hour by-12 \fi
	\ifnum\hour=0 \hour=12 \fi
	\number\hour:\ifnum\minute<10 0\fi\number\minute\the\amorpm}}
\edef\militarytime{\number\hour:\ifnum\minute<10 0\fi\number\minute}

\def\draftlabel#1{{\@bsphack\if@filesw {\let\thepage\relax
   \xdef\@gtempa{\write\@auxout{\string
      \newlabel{#1}{{\@currentlabel}{\thepage}}}}}\@gtempa
   \if@nobreak \ifvmode\nobreak\fi\fi\fi\@esphack}
        \gdef\@eqnlabel{#1}}
\def\@eqnlabel{}
\def\@vacuum{}
\def\marginnote#1{}
\def\draftmarginnote#1{\marginpar{\raggedright\scriptsize\tt#1}}
\def\draft{
	\pagestyle{plain}
	\overfullrule=2pt
        \oddsidemargin -.5truein
        \def\@oddhead{\sl \phantom{\today\quad\militarytime} \hfil
        \smash{\Large\sl DRAFT} \hfil \today\quad\militarytime}
        \let\@evenhead\@oddhead
        \let\label=\draftlabel
        \let\marginnote=\draftmarginnote
        \def\ps@empty{\let\@mkboth\@gobbletwo
        \def\@oddfoot{\hfil \smash{\Large\sl DRAFT} \hfil}
        \let\@evenfoot\@oddhead}
        \def\@eqnnum{(\theequation)\rlap{\kern\marginparsep\tt\@eqnlabel}%
        \global\let\@eqnlabel\@vacuum}  }

\catcode`\@=12

\input epsf
\def\cpsbox{\epsfcheck\cpsbox}
\def\epsfcheck{\ifx\epsfbox\UnDeFiNeD
	\message{(NO epsf.tex, FIGURES WILL BE IGNORED)}
	\gdef\cpsbox##1##2##3##4##5{\vbox to ##2
                {\hbox to ##1 {\hss} \vss}}
\else\gdef\cpsbox##1##2##3##4##5{
	\centerline{\epsfxsize=##1
\epsfysize=##2
\epsfbox{##3}}
	\nobreak
	\centerline{\epsfxsize=##1
\epsfysize=##2
\epsfbox{##4}}
	\nobreak
	\centerline{\epsfxsize=##1
\epsfysize=##2
\epsfbox{##5}}
	}\fi}
\def\psinsert#1#2#3#4#5#6{
	\goodbreak\medskip
  	\cpsbox{#1 \hsize}{#2}{#3}{#4}{#5}
	\nobreak\medskip
	\centerline{
	   \vbox{
	   \hsize=.9\hsize
	   \small \def\baselinestretch{1.1} #6 } }
\bigskip\goodbreak
}

\def\lae{\smash{\,\lower .5 ex \hbox{$\,\stackrel<\sim\,$}}}
\def\gae{\smash{\,\lower .5 ex \hbox{$\,\stackrel>\sim\,$}}}

\def\L{{\cal L}}

\def\beq{\begin{equation}}
\def\eeq{\end{equation}}

\def\sutw{${\rm SU}(2)_W$}

\def\KKb{$K^0-\bar K^0$}
\def\DDb{$D^0-\bar D^0$}

\def\Kpinunu{$K^+\longrightarrow\pi^+\nu\bar\nu$}
\def\Kee{$K_L\longrightarrow e^+e^-$}

\begin{document}
\setlength{\baselineskip}{3.0ex}

\hfill    WIS--93/89/Sept--PH
\vspace*{2.3cm}
\begin{center}
{\large\bf New Bounds on Leptoquarks}\\
\vspace*{6.0ex}
{\large Miriam Leurer} \\
\vspace*{1.5ex}
{\large\it Department of Nuclear Physics\\
The Weizmann Institute\\
Rehovot 76100\\
ISRAEL}
\vspace*{1.5ex}
\vspace*{3.0ex}
\vspace*{6.0ex}
\end{center}
\vspace*{6.0ex}
\begin{center}
{\bf Abstract}
\end{center}

\noindent
We show that FCNC processes are unavoidable for leptoquarks that couple to left
handed quarks, and derive new FCNC bounds from neutral meson mixing. Despite
being induced only at one loop, these processes lead to significant bounds
since the leptoquark contributions do not suffer from GIM cancellations.
Studying the implications of these bounds we find that (i) The \DDb{} mixing
bound is the first significant FCNC bound from the up sector. Combining it with
FCNC bounds from the down sector we arrive at a bound on the first generation
couplings. (ii) The \KKb{} and \DDb{} mixings bound $g^2/M$ while all other
processes bound $g/M$. The combined neutral meson mixing bound is therefore
dominant for the heavier leptoquarks, and leads to exclusion of large regions
in parameter space which were previously allowed.

\begin{center}
{\noindent \it (Contributed talk at the
5th International Symposium on Heavy Flavour Physics,\\
McGill University, July 6-10, 1993)}
\end{center}

\newpage
It is well known that \KKb{} mixing is a sensitive probe for beyond-standard
models \cite{BoHa}. In this talk we will show that \DDb{} mixing has become a
probe almost as sensitive as that of \KKb{} mixing.

We will study here the bounds on light scalar leptoquarks. Since for many of us
leptoquarks are associated with the GUT scale, we  will start by explaining how
can a leptoquark be light. Basically, there are three conditions that light
leptoquarks should obey: They should {\it not} couple to diquarks, and they
should couple chirally and diagonally. We will now explain in some detail the
meaning of these conditions:\newline
$\bullet$ Diquark couplings are forbidden, since they, together with the
lepton-quark
couplings lead to nucleon decay. The bound on the leptoquark mass
is then extremely strong, of the order of the scale of grand-unified theories.
\newline
$\bullet$ When we say that a leptoquark couples chirally, we mean that it
couples {\it either} to left--handed (LH) {\it or} to right--handed (RH)
quarks, not to both. A nonchiral leptoquark induces the following four-Fermi
interaction:
\beq
\L_{eff}=\frac{g_L g_R}{2M^2}~\bar u_R d_L~\bar e_R \nu_L
\label{gLgR}\;
\eeq
Here $g_L$ and $g_R$ are the leptoquark couplings to LH and RH quarks
respectively and $M$ is its mass. The above interaction contributes to
$\pi\longrightarrow e\nu$ decay and, in contrast to the standard model
interaction, it is not chiral and its amplitude is not helicity suppressed. The
amplitude is therefore enhanced by $m_\pi/m_e$ relative to the standard model
amplitude and, in addition, it is possible to show that there is further
enhancement by $m_\pi/(m_u+m_d)$ \cite{Shanker}. The enhanced effect of the
interaction (\ref{gLgR}) leads to unacceptable deviations from lepton
universality in $\pi$ decays, unless one strongly constrains the leptoquark
parameters with the 95\% CL bound as strong as $M^2/g_L g_R\geq (170~{\rm
TeV})^2$. The chirality requirement enables us to circumvent this bound.
\newline $\bullet$ Leptoquarks couplings are called ``diagonal'' when the
leptoquark couples to a single leptonic generation and to a single quark
generation. If the leptoquarks couple nondiagonally they induce flavour
changing neutral current (FCNC) processes in both the leptonic sector and the
quark sector, leading to strict bounds on the leptoquark parameters
\cite{PatiS}, \cite{Buch1}. To avoid these bounds we impose diagonality of the
couplings.

The first important point we want to make here is that diagonality is not
really possible for leptoquarks that couple to left-handed quarks. The fact
that the CKM matrix \cite{CKM} is not trivial implies that one cannot
diagonalize the leptoquark interactions simultaneously in the up and the down
quark sectors. For example, if the couplings to the up sector are diagonal, and
the leptoquark couples only to the first generation up quark, then the
couplings in the down sector are not diagonal: The leptoquark couples
``mainly'' to the down quark, but there are also some coupling to the strange
quark (suppressed by $\sin\theta_C$) and some coupling to the bottom quark
(suppressed by $V_{13}$, where $V$ is the CKM matrix). Similarly, if the
leptoquark couplings to the down quark are diagonal, then its couplings to the
up quark sector are {\it almost} diagonal, but not strictly so. Therefore FCNC
processes are unavoidable for leptoquarks that couple to LH quarks. We will now
examine these leptoquarks more closely and will strive to make their
couplings as close to diagonal as possible, so that the bounds from FCNC
processes will be minimal. These minimal FCNC bounds are {\it utterly
unavoidable} and turn out to be very significant for heavier leptoquarks (above
a few hundred GeV) \cite{me}.

There are three scalar leptoquarks that can couple to LH quarks, $S$, an
\sutw{} scalar carrying weak hypercharge $Y=1/3$; $D$, an \sutw{} doublet with
$Y=-7/6$; and $T$, an \sutw{} triplet with $Y=1/3$. Their couplings are given
by:
\begin{eqnarray}
\L_{S} &=& \sum_i \, \left(g_i~\bar e^c u^i_L - g_i'~\bar\nu^c d^i_L \right)
               \, S^{(1/3)}\nonumber\\
\L_{D} &=& \sum_i \, \left\{g_i~\bar e\, u^i_L D^{(-5/3)} +
                            g_i'~\bar e\, d^i_L D^{(-2/3)} \right\} \nonumber\\
\L_T &=& \sum_i \, \left\{{\sqrt2} g_i~\bar \nu^c u^i_L T^{(-2/3)}
      + (g_i~\bar e^c u^i_L + g_i'~\bar\nu^c d^i_L) \, T^{(1/3)}
      + {\sqrt2} g_i'~\bar e^c d^i_L T^{(4/3)} \right\}  \; ,
\label{YukawaL}
\end{eqnarray}
where the $g_i$'s and $g_i'$'s are related by a CKM rotation: $g_i'=g_j
V_{ji}$, with $V$ the CKM mixing matrix. In  order to present our bounds we
define the overall strength of the Yukawa couplings to be $g$ with $
g=\sqrt{\sum_i~|g_i|^2} $, and give our final results as bounds in the $g$ --
$M$ plane.

Since leptoquarks that couple to the first generation, are currently of the
most interest \cite{HERA}, we will assume from now on that the leptoquarks $S$,
$D$ and
$T$ couple to the first generation of leptons and ``mainly'' to the first
generation of quarks, that is, we will assume that the second generation
couplings are suppressed by $O(\sin\theta_c)$ and the third generation
couplings by $O(|V_{13}|+|V_{12}\cdot V_{23}|)$. Since the third
generation couplings are so suppressed, they will actually have no effect, so
we shall ignore them and reduce to a two generation picture.
Then the couplings can be parametrized by:
\begin{eqnarray}
g_1=g\cos\theta~~~~~&{\rm and}&~~~~~g_2=-g\sin\theta \nonumber \\
g'_1=g\cos(\theta_C-\theta)~~~~~&{\rm
and}&~~~~~g'_2=g\sin(\theta_C-\theta)
\label{gpar}\;.
\end{eqnarray}
The angle $\theta$ describes the deviation from diagonality in the up sector,
while $(\theta_C-\theta)$ describes the deviation from diagonality in the down
sector. $\theta$ therefore determines the division of the FCNC problems
between the two quark sectors. Our purpose is to find ``the best'' $\theta$,
the one which will soften all bounds as much as possible.

At this point we reach the main issue of this talk: {\it Up till now no
significant FCNC bounds were known to arise from the up sector}. The best
$\theta$ choice was obviously $\theta=\theta_C$, FCNC were then hidden in the
up sector, and could not provide any observable signature. However, we want to
show here that there is a new class of FCNC bounds, arising from neutral meson
mixings and including a significant bound on the up quark sector couplings. One
could, at first thought, discard neutral meson mixings as unimportant, since
leptoquarks induce them only at one loop, in contrast to other leptoquark
bounds that arise already at tree level. However, such an approach is mistaken:
After all, \KKb{} and \DDb{} mixing arise in the standard model too only at one
loop. Moreover, the GIM mechanism of the standard model leads to a suppression
of {\it e.g.} \KKb{} mixing by $(m_c/M_W)^2$, while for the leptoquarks
contribution
there is no suppression of this kind. We therefore should expect that neutral
meson mixing will supply us with significant bounds on the leptoquarks
parameters. Actually, the fact that the leptoquark contributions arise
only at the one loop level is in some ways advantageous: Tree level processes
induced by
the leptoquarks lead to bounds on $g/M$, while the one loop \KKb{} and \DDb{}
mixings lead to bounds on $g^2/M$. Therefore, the neutral meson mixing bounds
will always become the dominant bounds at some high mass region.

Table 1 summarizes our bounds on the $S$, $D$ and $T$ leptoquarks
that couple to the first generation of leptons and ``mainly'' to the first
generation of quarks. These bounds arise from universality in $\pi$ decay, from
atomic parity violation \cite{me2}, and from FCNC processes: Rare $K$ decays
and \KKb{} mixing in the down sector and \DDb{} mixing in the up sector. All
the bounds are at the 95\% CL. For a detailed derivation of these bounds see
\cite{me2}.
\begin{table}
\def\arraystretch{1.2}
\begin{center}
\begin{tabular}{|c|c|c|}\hline
$S$&$D$&$T$
\\ \hline
$3100 \, g$ &$1800 \, g$  & $1000 \, g$ \\
$\pi\longrightarrow e\nu$ & $\pi\longrightarrow e\nu$ & Atomic Parity \\ \hline
\def\arraystretch{1.1}
$\vphantom{\biggl|}
4700 \, \sqrt{{|g_1'g_2'|}/{\sin\theta_C}}$ &
$5950 \, \sqrt{{|g_1'g_2'|}/{\sin\theta_C}}$  &
$4200 \, \sqrt{{|g_1'g_2'|}/{\sin\theta_C}}$  \\
\Kpinunu&\Kee&\Kee \\ \hline
$\vphantom{\biggl|}8000 \, {{|g_1'g_2'|}/{\sin\theta_C}}$ &
$8000 \, {{|g_1'g_2'|}/{\sin\theta_C}}$ &
$18000 \, {{|g_1'g_2'|}/{\sin\theta_C}}$ \\
\KKb&\KKb&\KKb \\ \hline
$\vphantom{\biggl|}4500 \, {{|g_1g_2|}/{\sin\theta_C}}$ &
$4500 \, {{|g_1g_2|}/{\sin\theta_C}}$ &
$10000 \, {{|g_1g_2|}/{\sin\theta_C}}$ \\
\DDb & \DDb & \DDb \\
\hline
\end{tabular}
\end{center}
\null
\medskip
\centerline{\vbox{ \hsize=.9\hsize \small \def\baselinestretch{1.1}
	\bf Table 1. \it 95\% CL lower bounds on the leptoquark masses, in GeV,
	as a function of the coupling constants.}}
\end{table}

Our task is to combine the FCNC bounds from both sectors to a bound on the
overall coupling $g$. The bounds from the up sector apply to the
coupling constant combination $|g_1g_2|$, and those from the down sector to
$g_1'g_2'$, namely:
\begin{eqnarray}
f_u(M)&\geq|g_1g_2|&=g^2|\sin(2\theta)/2| \nonumber\\
f_d(M)&\geq|g_1'g_2'|& = g^2|\sin2(\theta_C-\theta)|/2
\label{gcombi}\;,
\end{eqnarray}
where $f_u(M)$ is the \DDb{} mixing bound and is linear in $M$, and $f_d(M)$ is
the dominant FCNC bound from the down sector: At low masses $f_d(M)$ arises
from rare $K$ decays and is then quadratic in $M$, at higher masses it comes
from \KKb{} mixing and so is also linear in $M$. Equation (\ref{gcombi}) makes
it clear that {\it any} choice of $\theta$ will lead to bounds on $g^2$. The
``best'' angle $\theta(M)$ that leads to the softest bounds, is given by
saturating both inequalities in (\ref{gcombi}), then:
\beq
\frac{f_u(M)}{f_d(M)}=\left|\frac{\sin2\theta}{\sin2(\theta_C-\theta)}\right|
\label{theta}\;.
\eeq
Solving equation (\ref{theta}) for the ``best'' angle $\theta$, and
substituting this angle into any of the two inequalities of (\ref{gcombi})
we get a bound on the overall coupling $g$. Since $\theta$ is small
($\theta\leq \frac{2}{3}\theta_C$), the overall coupling is equal, to a very
good approximation, to $g_1$ and $g_1'$. {\it We therefore interpret the bound
on $g$ as a bound on the couplings to the first generation}.

\goodbreak
Figures 1(a--c) describe the combined FCNC bounds for each of the leptoquarks
$S$, $D$ and $T$ and compare them to the conventional bound (in the first row
of table 1). The FCNC bounds become stronger than the conventional bound at
$M=3300~$GeV for $S$, at $M=270~$GeV for $D$ and at $M=710~$GeV for the $T$
leptoquark.

\psinsert{.9}{2.1in}{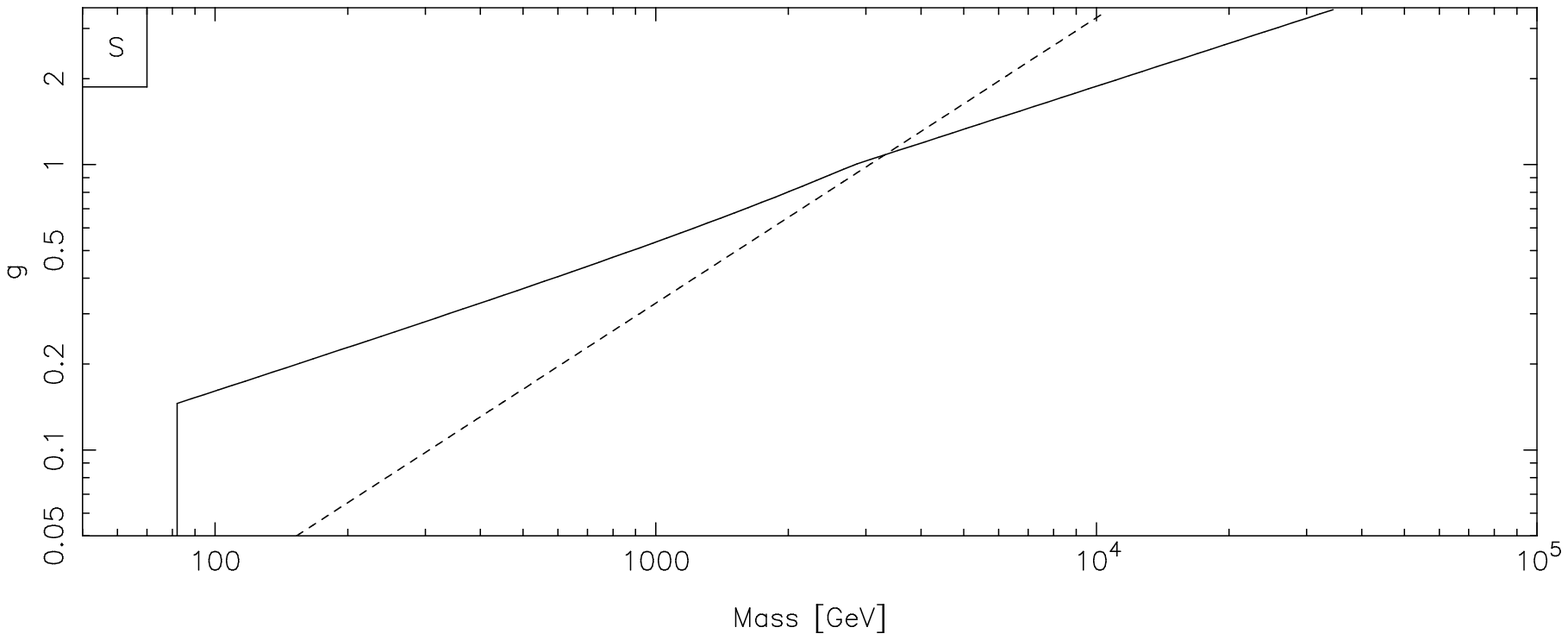}{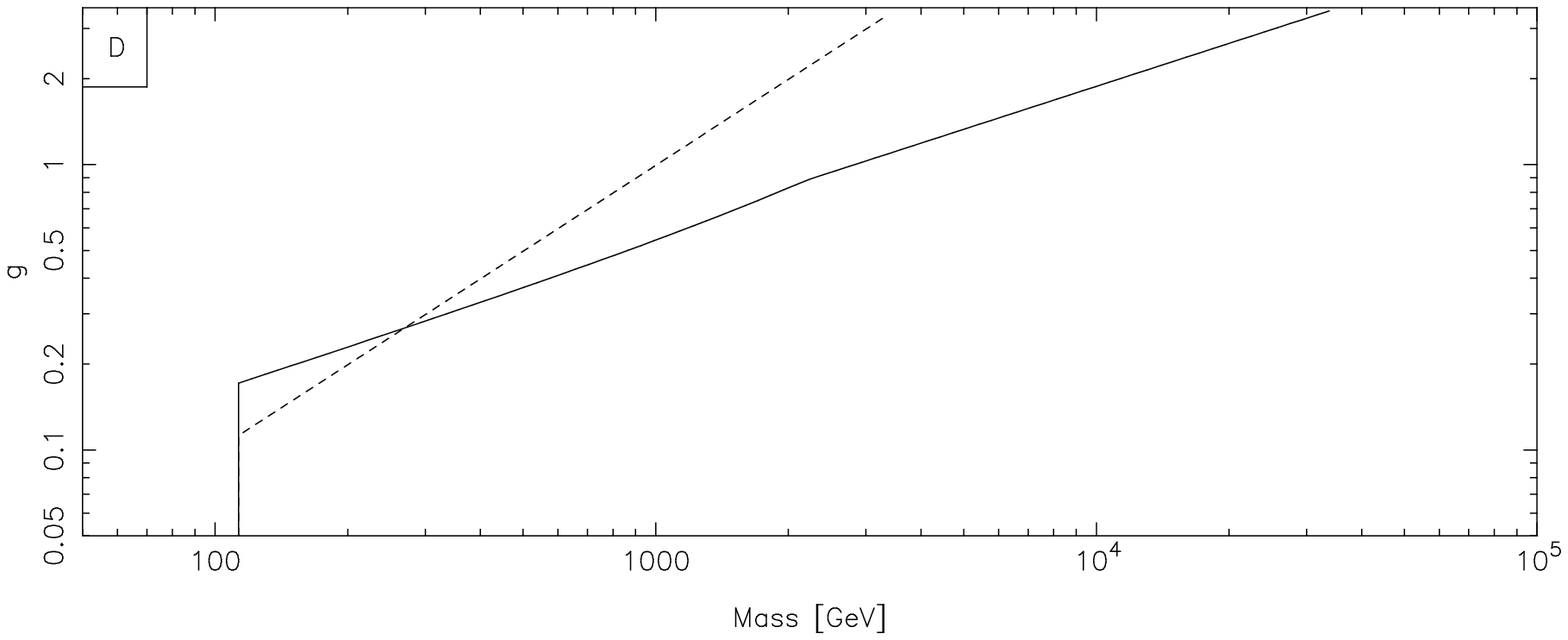}{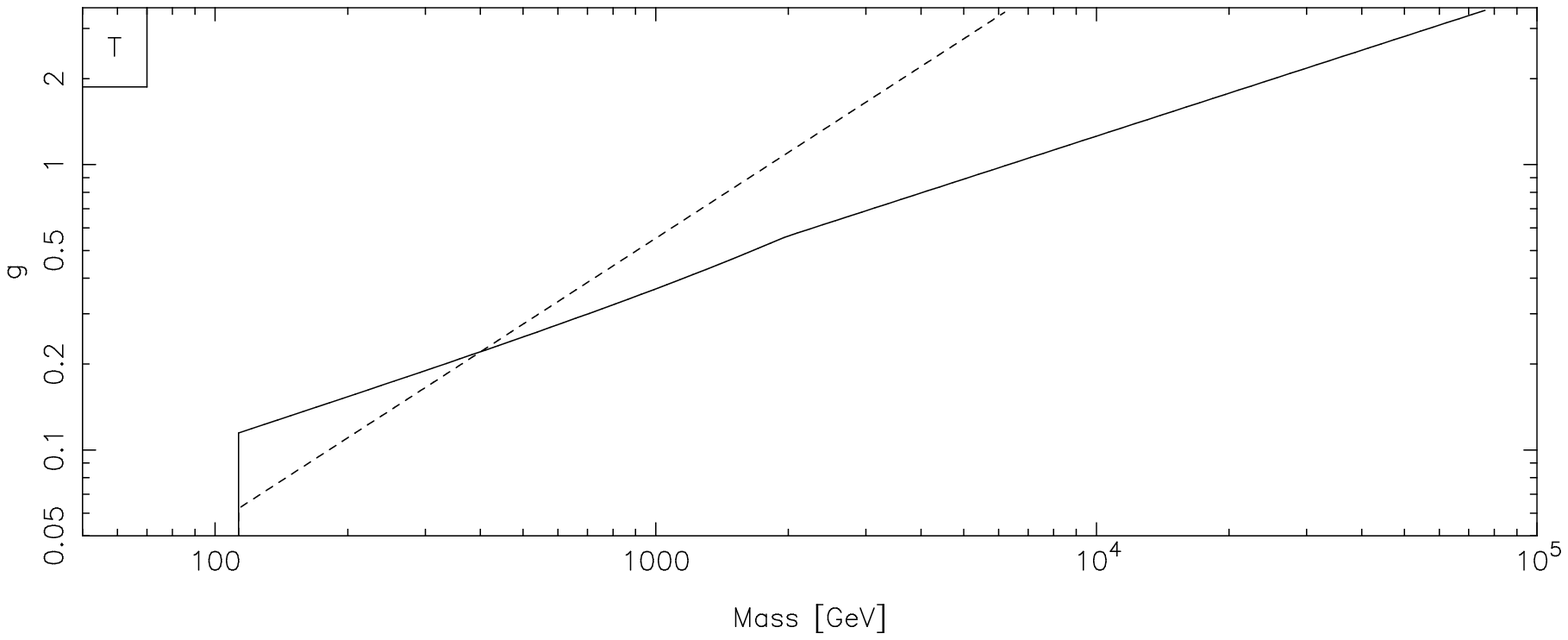}{
\noindent {\bf Figure 1(a--c).} {\it
Bounds on the couplings and masses of first generation leptoquarks that couple
to LH quarks, $S$, $D$ and $T$. The regions above the lines are excluded. The
full line in each figure is our new FCNC bound, the dashed line is the
``conventional'' bound arising from universality in leptonic $\pi$ decays or
atomic parity violation. We took the direct CDF bounds \cite{CDF} into account
and therefore the mass ranges are above 82$~$GeV for $S$ and above 113$~$GeV
for
$D$ and $T$.}}

Our new bounds are important for various proposals for leptoquark searches,
particularly for {\it indirect} searches in accelerators \cite{proposals}: We
exclude leptoquarks with coupling $g=e$ (where $e$ is the electromagnetic
coupling) at masses $930~$GeV, $340~$GeV and $710~$GeV for $S$, $D$ and $T$,
respectively. Comparing these results with recently proposed methods
\cite{proposals} for indirect leptoquark searches we find that our bounds
already exclude large parts of the parameter space that could be penetrated by
these methods. Our results are also interesting for the direct searches at
HERA: At the moment, our bounds are better than the first results from HERA
for all masses, but in the future HERA will improve on our bounds at
masses below $\sim 300~$GeV.
\vskip 0.8 cm
\noindent
{\bf Acknowledgement:} I thank Neil Marcus for many helpful discussions.


\begin{thebibliography}{9}
\bibitem{BoHa} R.N.~Cahn and H.~Harari, \np176,135,80.
\bibitem{Shanker} O.~Shanker, \np204,375,82.
\bibitem{PatiS} J.C.~Pati and A.~Salam \pr10,275,74.
\bibitem{Buch1} W.~Buchm\"uller and D.~Wyler, \pl177,377,86.
\bibitem{CKM} N.~Cabibbo, \prl10,531,63; M.~Kobayashi and T.~Maskawa,
\pro49,652,73.
\bibitem{me} M.~Leurer, \prl71,1324,93.
\bibitem{HERA} M.~Derrick {\it et al.\/} (ZEUS Collab.), \pl 306,173,93;
I.~Abt {\it et al.\/}, (H1 Collab.) \np396,3,93.
\bibitem{me2} M.~Leurer, ``A comprehensive study of leptoquark bounds'',
Weizmann Institute preprint WIS--93/90/Sept--PH, hep-ph/9309266.
\bibitem{CDF} S.M.~Moulding, CDF Collab., Proc. Seventh Meeting of the APS,
Fermilab, USA (Nov. 1992), Fermilab preprint CONF-92-341-E.
\bibitem{proposals} An incomplete list includes:
O.J.P.~Eboli and A.V.~Olinto,\pr38,3461,88;
M.A.~Donchesky and J.L.~Hewett, \zp56,209,92;
H.~Nadeau and D.~London, \pr47,3742,93;
O.J.~ Eboli {\it et al.}, \pl311,147,93;
G.~B\'elanger, D.~London and H.~ Nadeau, ``Single leptoquark production at
$e^+e^-$ and $\gamma\gamma$ colliders'', hep-ph/9307324.
\end{thebibliography}
\end{document}